\documentclass{article}

\usepackage{arxiv}

\usepackage[utf8]{inputenc} 
\usepackage[T1]{fontenc}    
\usepackage{hyperref}       
\usepackage{url}            
\usepackage{booktabs}       
\usepackage{amsfonts}       
\usepackage{nicefrac}       
\usepackage{microtype}      
\usepackage{lipsum}
\usepackage{graphicx}
\graphicspath{ {./images/} }

\usepackage{colortbl}
\usepackage{bbm}
\usepackage{smartdiagram}
\usepackage{float}
\usepackage{here}
\usepackage{adjustbox}
\usepackage{booktabs}
\usepackage{amsmath}
\usepackage{tabularx}
\usepackage{enumitem}
\usepackage{hyperref}
\usepackage{capt-of}
\usepackage{listings}

\def\bSig\mathbf{\Sigma}

\title{Multilevel Conditional Autoregressive models for
longitudinal and spatially referenced epidemiological data}

\author{
Dany Djeudeu \\
  Faculty of Statistics\\ TU Dortmund University\\
   44221 Dortmund\\
  \texttt{deudjui@statistik.tu-dortmund.de} \\
   \And
 Susanne Moebus \\
  Institute for Urban Public Health\\
  University Hospital Essen\\
  45122 Essen \\
  \texttt{susanne.moebus@uk-essen.de} \\
  \And
 Katja Ickstadt \\
  Faculty of Statistics\\ TU Dortmund University\\
   44221 Dortmund\\
  \texttt{ickstadt@statistik.tu-dortmund.de} \\
}

\begin{document}
\maketitle
\begin{abstract}
The classical multilevel model fails to capture the proximity effect in epidemiological studies, where
subjects are nested within geographical units. Multilevel Conditional Autoregressive models are
alternatives to help explain the spatial effect better. They have been developed for cross-sectional
studies but not for longitudinal studies so far. This paper has two goals. Firstly, it further develops
the multilevel (growth) models for longitudinal data by adding existing area level random effect terms
with CAR prior specification, whose structure is changing over time. We name these models MLM
tCARs for longitudinal data. We compare the developed MLM tCARs to the classical multilevel
growth model via simulation studies in common spatial data situations. The results indicate the
better performance of the MLM tCARs, to retrieve the true regression coefficients and with better
fit in general.

Secondly, this paper provides a comprehensive decision tree for analysing data in epidemiological
studies with spatially nested structure: we also consider the Multilevel Conditional Autoregressive models
for cross-sectional studies (MLM CARs). We compare three models (for cross-sectional studies) via
simulation studies: the classical multilevel model, the multilevel CAR model and the Restricted
CAR model that accounts for spatial confounding. The MLM CARs, particularly the Restricted
CAR show better results.
We apply the models comparatively on the analysis of the association between greenness and
depressive symptoms in the longitudinal Heinz Nixdorf Recall Study. The results show negative
association between greenness and depression and a decreasing linear individual time trend for all
models. We observe very weak spatial variation and moderate temporal autocorrelation.
\end{abstract}


\section{Introduction}
\label{introduction}

Several issues arise when analysing public health data with participants nested within spatial
areas, three of which we discuss here. The first is that risk factors and covariates related to individual
level health outcome, may be given on different spatial resolutions. It is better to use all data on
their finest level given  in order to avoid the ecological and atomic fallacy (see, e.g. \cite{ecological_multilevel}, \cite{lost_aggre}, \cite{aggr_level}).
The second issue concerns the spatial effect.
Two main types  should  be considered: spatial autocorrelation and spatial heterogeneity. Spatial autocorrelation is closely related to the first law of geography  (\cite{geography_law}, \cite{spatialwall})
  and can arise for a number of reasons, for instance, due to  unmeasured/unavailable confounders. The second type  is the  grouping effect or spatial heterogeneity (\cite{context_heterogeneity}). Sometimes,  the  difference between the two types of spatial effects is not obvious.
The third issue concerns the dynamic effect for longitudinal data.
Not only do individuals  change over time, but (characteristics of) geographical units as well.  These changes are the consequences of unavailable/unmeasured area level covariates that may be changing over time and should, therefore, be considered (\cite{long_multilevel}, \cite{repeated_dynamic_group}).  Modelling and inference  should  exploit the dynamic  and the spatial  effect.

 Multilevel modeling (mostly 2-level) offers a framework to  take advantage of the hierarchical structure of the data (\cite{multi_public}, \cite{group_diff}) and is widely  used in many applications in the medical, educational  and social science (\cite{lit_multilevel_ed}, \cite{social_multilevel}, \cite{med_multi}, \cite{psycho_research}).  The classical multilevel model (MLM CL2) has multiple advantages,  but also  some limitations.
The inclusion of random effects  at the highest hierarchical  level  helps to adjust for  fixed effect estimates, for  missing or unavailable covariates with spatial structure.
  A central assumption of interest in the  MLM CL2 is that  these random effects  are mutually independent across spatial units,  modelling spatial heterogeneity only.
 In the Bayesian framework, this consists of assuming exchangeable unstructured  priors on  area level random effects. This is equivalent to a global smoothing towards the mean effect. However, for positive covariation between adjacent  units, positions of the spatial units are important; local smoothing using adjacency may be more appropriate.
 Figure \ref{adjacency} in the supplementary material shows the importance of accounting
for spatial autocorrelation instead of the spatial heterogeneity only.

Several works in the literature of health geography and spatial epidemiology, mostly for cross-sectional
analysis have recognised both the spatial heterogeneity and the spatial autocorrelation
and modified the MLM CL2 accordingly:  (\cite{browne_MLM}, \cite{hongwei_MLM}, \cite{autolog}). However, most  of these methods are neither useful when substantial autocorrelation is
expected nor do they exploit spatial adjacency. Therefore, structured priors  on the area level random effect  recognizing adjacency are of interest.
Markov Random Field models, particularly the Conditional Autoregressive (CAR) models (\cite{CARspatial}, \cite{hoef_CAR}), are  suitable for this task.
CAR priors   have the advantage of facilitating random effects
analysis under a Markov Chain Monte Carlo (MCMC) sampling approach.
Recent  works that combined the advantages of multilevel models and Markov Random Field for cross-sectional data  include  \cite{multilevel_interat1}, \cite{multilevel_interat2}, \cite{multilevel_interat3} and  the S.CARmultilevel() function of the CARBayes package (\cite{carspatialonly}).
 However,  spatial confounding (\cite{fixed_eff_you_love1}, \cite{fixed_eff_you_love2}) was not explicitly examined  in the previous  papers. We consider the multilevel model with restricted CAR model (MLM RCAR) to account for this spatial confounding.

 To examine the  change over time of the health outcome in longitudinal studies, classical  $3$-level growth models (also MLM CL3) are widely used when participants are nested within geographical units.  This consists of including area-level   random effects that are independent across spatial units in the model equation, to account for the spatial effect.
 In addition to aforementioned limitations of
MLM CL2 for cross-sectional analyses, the geographical units are conceived here as entities that
exert an effect that changes systematically with time. It is more realistic to assume that areas undergo structural and functional changes over time that are more stochastic in nature. These
changes are not well captured by the classical model MLM CL3 (\cite{repeated_dynamic_group}).

The contribution of this paper is twofold: The first and main objective is to further develop
Multilevel Conditional Autoregressive models for longitudinal data (MLM tCARs) by combining
some already existing models: The classical 3-level growth model and some CAR models to account
for spatio-temporal random effects. We compare the developed MLM tCARs in the longitudinal
setting to the classical (3-level) multilevel growth model (MLM CL3) in terms of accuracy and
stability in the coefficient estimates under the presence and absence of spatial effects in data via
a simulation study. The second goal is to provide epidemiologists with a practical decision tree
to help choose the appropriate models that produce more stable coefficient estimates in case of
nested data, for cross-sectional as well as longitudinal studies. To achieve this, we also compare the
MLM CARs for cross-sectional data to the MLM CL2 in a simulation study. In contrast to existing
studies (\cite{multilevel_interat3}),we additionally analyse whether adding the spatially correlated error
term (CAR-prior term) to a linear model shrinks or enlarges/inflates the true regression coefficients.
The Restricted Multilevel Conditional Autoregressive Model (MLM RCAR) (\cite{restricted_CAR}) accounts for spatial confounding. Thus, we compare the MLM RCAR, the MLM CAR and the
MLM CL2 for cross-sectional data.

We organized our article as follows: In the first part of section \ref{methode} we shortly present the MLM CL2 and  MLM CARs for cross-sectional studies. In the second part of section \ref{methode}, we introduce
the developed MLM tCARs for longitudinal data and then present selected comparison criterion
 for epidemiological and public health applications. The simulation strategy follows. The method
section ends with a computation subsection, which summarizes the techniques used for simulation
and model fitting. In section \ref{results} we outline the results of the simulation studies.  The decision tree for
the most appropriate methods,  when spatial effects are involved, is provided in section \ref{guideline}.
 In section  \ref{application} we apply the MLM CL3 and MLM tCARs for longitudinal studies comparatively to the analysis
 of the association between depression and greenness in the longitudinal Heinz Nixdorf Recall Study (HNRS).  Section \ref{discussion} is dedicated to the discussion of the results  of the simulation studies  and the application.

\section{Methods}\label{methode}

\subsection{Multilevel Conditional Autoregressive models for cross-sectional data}\label{carcross}
For multilevel models for cross-sectional data, the study region is partitioned into $K$ non-overlapping areal units. Data are available on   $n=\sum_{j=1}^{K}n_{j}$ individuals, with $n_{j}$ individuals within area $j$, $j=1, \ldots, K$.
 The following model  equation accounts for both spatial heterogeneity and spatial autocorrelation:
\begin{equation}\label{two_level}
\left\{
  \begin{array}{ll}
     y_{ij} &= X_{ij}^{T}\beta + \psi_{j} + e_{ij},\\
  \psi_{j}|\psi_{-j}, W &\sim N\left(\displaystyle\frac{\rho\sum\limits_{k \neq j}w_{jk}\psi_{k}}{\rho\sum\limits_{k \neq j}w_{jk} +1-\rho} \displaystyle\frac{\tau^{2}}{\rho\sum\limits_{k \neq j}w_{jk} +1-\rho}\right),\\
  e_{ij} &\sim N(0, \sigma^{2}_{e}) ~~\forall~~  i,j, ~~ i=1, \ldots n,~~ j=1,\ldots, K,
  \end{array}
\right.
\end{equation}
where $\psi_{-j} = (\psi_{1}, \psi_{2}, \ldots, \psi_{j-1}, \psi_{j+1}, \ldots, \psi_{K})$, $\sigma^{2}_{e}, \tau^{2}$ follow an inverse Gamma $IG(a,b)$ distribution, $\rho$ follows a uniform distribution $R(0,1)$.
 Here, $X_{ij}^{T}$  is  a $ 1\times p$ vector of  intercept and $p-1$ covariates for individual $i$ in area $j$.
 $X_{ij}^{T}$ includes individual level as well as area level covariates. $W = (w_{kj})_{k,j = 1\ldots K}$ is the binary adjacent matrix.
The area level random effect vector $(\psi_{1}, \ldots, \psi_{K})^{T}$ has the \textbf{Leroux} structure given in equation (\ref{two_level}) (\cite{leroux}, \cite{congdon_hier}, p. $181-183$).
$\rho=0$ corresponds to a lack of spatial interdependence, i. e.  the classical multilevel model (2 levels, MLM CL2).
By contrast, $\rho=1$ leads to the intrinsic CAR  (\cite{congdon_hier}, p. $183-184$) model.

Spatial confounding can be interpreted in linear-model terms as a collinearity problem.
When spatial confounding is detected,  spatial smoothing  is restricted to the orthogonal complement of the fixed effect of area level variables, called Restricted Conditional Autoregressive model (RCAR). This is recommended and described in \cite{fixed_eff_you_love1}, \cite{fixed_eff_you_love2}. Choosing the
restricted CAR models for the area level random effect instead of the CAR models leads to the
MLM RCAR.

\subsection{Multilevel Conditional Autoregressive models for longitudinal data}
Our model combines the advantages of multilevel models and the properties of the Markov Random Field to accurately model the random effects in a multilevel growth model. It is defined as follows:

\begin{equation}\label{MLMCAR}
\left\{
  \begin{array}{ll}
   y_{tij} &= X_{tij}^{T}\beta + \psi_{tj} + r_{0ij} + r_{1ij}g(t) + e_{tij},\\
  \psi_{tj} &\sim  \text{(refer to equations (\ref{caranova_eq}) and (\ref{conv_rand}) for candidate models),}\\
  (r_{0ij}, r_{1ij})^{T} &\sim N\left(\left(
                                         \begin{array}{c}
                                           0 \\
                                           0 \\
                                         \end{array}
                                       \right)
  , \left(
                                  \begin{array}{cc}
                                   \sigma_{r_{0}}  & \sigma_{r_{01}} \\
                                    \sigma_{r_{01}} & \sigma_{r_{1}} \\
                                  \end{array}
                                \right)
  \right)\\
       e_{tij} &\sim N(0, \sigma^{2}_{e}) ~~\forall~~ t,i,j, ~~~ t=1, \ldots,N, ~~ i=1, \ldots n,~~ j=1,\ldots, K.
  \end{array}
\right.
\end{equation}

$y_{tij}$ is a continuous outcome for individual $i$ in the spatial unit $j$ at the measurement occasion $t$.
 $X_{tij}^{T}$ is  a $ 1\times p$ vector of intercept and $p-1$ covariates for individual $i$ in area $j$ at the measurement occasion $t$.
 $X_{tij}^{T}$ includes individual level and time-varying variables, individual level and time-invariant variables,  area level and time-varying variables, and area level and time-invariant variables. It also includes a  deterministic function of time $g$, defining the individual growth. Note that g could be defined differently for each individual. $\beta = (\beta_{0}, \ldots, \beta_{p-1})^{T}$ with prior $\beta \sim N(0, \Sigma_{\beta})$ is the vector of  regression coefficients.
  There are three levels represented by sources of random variation at the area level, the individual level and the observation level in the random effect part.  We assume the outcome for individuals $i$  in
spatial areas $j$ and time $t$ are conditionally
independent and normally distributed. $r_{0ij}$ and $r_{1ij}$, independent of $e_{tij}$, are random effects for the  individuals to account for person-to-person differences in the repeated measures and how they change over time.
$(\psi_{t1},\ldots, \psi_{tK})$ is the vector of random effects  for time period $t$, which evolves over time and makes use of temporal and spatio-temporal dynamics.  Independent of $e_{tij}$,  $r_{0ij}$ and $r_{1ij}$,
it decomposes the spatial effects into spatial heterogeneity and spatial autocorrelation.
There are existing models for $\psi_{tj}$ in the literature with this structure, mostly used in disease mapping models so far, including  (\cite{Knorr_phi}, \cite{Lee_phi}, \cite{Bernardinelli_phi}).
Such models also help to understand the dynamic of the spatial effect. We select three models for $\psi_{tj}$ for our model in equation (\ref{MLMCAR}) and
for our simulation study: the CAR ANOVA model, the convolution model, and the classical model.
This lead to the MLM CAR ANOVA, MLM CONV und MLM CL3, respectively. Together, the MLM CAR ANOVA and MLM CONV are called MLM tCARs.

The CAR ANOVA model for $\psi$ is given by

\begin{equation}\label{caranova_eq}
\left\{
  \begin{array}{ll}
   \psi_{tj} &= \phi_{j}  +\delta_{t} + \omega_{tj}, \\
     \phi_{j} |  \phi_{-j}, W  &\sim N\left( \displaystyle\frac{\rho_{S}\sum\limits_{k \neq j}w_{jk}\phi_{k}}{\rho_{S}\sum\limits_{k \neq j}w_{jk}+1-\rho_{S}}, \displaystyle\frac{\tau^{2}_{S}}{\rho_{S}\sum\limits_{k \neq j}w_{jk}+1-\rho_{S}}\right), \\
      \delta_{t} |  \delta_{-t}, D  &\sim N\left(\displaystyle\frac{\rho_{T}\sum\limits_{l \neq t}d_{tl}\delta_{l}}{\rho_{T}\sum\limits_{l \neq t}d_{tl}+1-\rho_{T}}, \displaystyle\frac{\tau^{2}_{T}}{\rho_{T}\sum\limits_{l \neq t}d_{tl}+1-\rho_{T}}\right), \\
       \omega_{tj} & \sim N(0, \sigma_{\omega}^{2}), ~~~ t=1, \ldots,N, ~~ j=1,\ldots, K,\\
    \end{array}
\right.
\end{equation}
where $\phi_{-j} = (\phi_{1}, \phi_{2}, \ldots, \phi_{j-1}, \phi_{j+1}, \ldots, \phi_{K})$,  $\delta_{-t} = (\delta_{1}, \delta_{2}, \ldots, \delta_{t-1}, \delta_{t+1}, \ldots, \delta_{T})$.
 $\rho_{S}, \rho_{T}  \sim R(0,1)$, as well as $\tau^{2}_{S}, \tau^{2}_{T}, \sigma_{\omega}^{2} \sim IG(a, b)$, see \cite{Knorr_phi}.
The model decomposes the
spatio-temporal variation in the data into $3$ components: an overall spatial effect common
to all time periods, an overall temporal trend common to all spatial units, and a set of independent space-time interactions. This is an ANOVA-type decomposition.  This model is appropriate if the goal is to estimate
overall time trends and spatial patterns.
Here, the spatio-temporal autocorrelation is modelled by a common set of spatial random
effects $\phi = (\phi_{1}, \ldots, \phi_{K})$  and a common set of temporal random effects
$\delta = (\delta_{1}, \ldots, \delta_{T})$.
Both are modelled by the CAR prior proposed by \cite{leroux}.  $W$ is the same as in  subsection \ref{carcross}
while $D=(d_{tj})$ is the binary $N \times N$ temporal
adjacency matrix defined by $d_{tl}=1$ if $|l-t|=1$  and  $d_{tl}=0$ otherwise, $t, l = 1,\ldots, N$.
Additionally, the model
can incorporate an optional set of independent space-time interactions $\omega = (\omega_{11}, \ldots, \omega_{NK})$. For the inverse Gamma prior for the variance components  $IG(a,b)$, values for $a$ and $b$ could be $a = 1$, $b = 0.01$.

The convolution model for $\psi$ is defined by
\begin{equation}\label{conv_rand}
\left\{
  \begin{array}{ll}
  \psi_{tj} &= \phi_{tj} + \omega_{tj}, \\
    \phi_{tj}| \phi_{-tj}, W &\sim N\left(\bar{\phi}_{tj}, \sigma_{\phi_{tj}}^{2}\right), ~~~~ \bar{\phi}_{tj}  = \displaystyle\frac{\sum\limits_{k \neq j}w_{jk}\phi_{tk}}{\sum\limits_{k \neq j}w_{jk}}, ~~~~  \sigma_{\phi_{tj}}^{2} =  \displaystyle\frac{\tau^{2}_{t}}{\sum\limits_{k \neq j}w_{jk}},\\
    \omega_{tj} & \sim N\left(0, \sigma_{\omega t}^{2}\right), ~~~ t=1, \ldots,N, ~~ j=1,\ldots, K,
  \end{array}
\right.
\end{equation}
where $\tau_{t}^{2}$, $\sigma_{\omega t}^{2}$ $\sim$ $IG(a,b)$.
$\phi_{-tj}  = (\phi_{t1}, \ldots,\phi_{t(j-1)}, \phi_{t(j+1)}, \ldots, \phi_{tK})$. It
 is  an extension of the model originally by  \cite{CARspatial}. The spatial autocorrelation parameter $\tau_{t}^{2}$ as well as the spatial heterogeneity parameter $\sigma_{\omega t}^{2}$  are allowed to vary over time. i.e., the model produces
a separate effect for each area and each time point.

The Classical linear growth model
is the (3-level) model for which $\psi_{tj}=u_{0j} + g(t)u_{1j}$, reducing to $\psi_{tj}=u_{0j}$, depending on the goal of the analysis. Here, $u_{0j} \sim N(0, \sigma^{2}_{u_{0}})$ and $u_{1j} \sim N(0, \sigma^{2}_{u_{1}})$ capture the area level random effect or unexplained spatial variation for the intercept and slope respectively.

\subsection{Model comparison criterion for epidemiological studies}\label{compare_section}

Since we are mostly interested in association-based (regression) models applied to observational
data, we concentrate on comparison methods for which uncertainty in coefficient estimates are
examined.

 Let  $\hat{\beta}_{1}^{M}, \ldots, \hat{\beta}_{p}^{M}$ and  $\hat{sd}_{1}^{M}, \ldots, \hat{sd}_{p}^{M}$ be the estimated regression coefficients  and standard deviations respectively,  using regression model $M$ to fit the generated data.
The bias for the coefficient $i$, using model $M$ is defined by  $|\hat{\beta}_{i}^{M}-\beta_{i}|$, where $\beta_{1}, \ldots, \beta_{p}$ are the true regression coefficients.
The
Root Mean Square Error $RMSE(\hat{\beta}_{i}^{M})  = \displaystyle\sqrt{(Bias(\beta_{i}^{M}))^{2} + Var(\hat{\beta}_{i}^{M})}$ assesses the quality of the estimators of the true regression coefficients using the underlying model $M$.

We use the Deviance Information Criterion (DIC)  (\cite{DIC_selection}) as an in-sample prediction criterion, which is a reasonable choice for our  nested models. When comparing candidate models, smaller values of DIC indicate  better models.
 The DIC should be used together with the posterior log-likelihood  before recommendation. The model maximizing the posterior log-likelihood is preferable.

\subsection{Simulation strategy}

The simulation study is motivated by data situations typically observed in spatial epidemiology or
health geography, where data are collected on different spatial levels. The main goal of the simulation
study is to examine, how well the true regression coefficients for the candidate models are retrieved
for simulated spatial effects. We use the geography of the HNRS.

We start with the simulation
study for longitudinal data, based on equation (\ref{MLMCAR}).
We simulate the spatio-temporal random effect $\psi$. Then, we simulate two covariates at the individual level, one of which is time-varying, from normal distributions. One time-varying variable is simulated at the area level, from a normal distribution.  We consider a linear individual time trend $g(t) = t$.
 We hold predictor variables  fixed as well as the individual level error term.
After simulating the area level spatio-temporal random effect $\psi_{tj}$  for each scenario,  we generate the dependent variable.  More details on the model equations
used for simulating the spatio-temporal random effect are given in section \ref{SM}, equation (\ref{psi_long}).
$\rho_{S}$ and $\rho_{T}$ are spatial and temporal autocorrelation parameters respectively, with values  in the unit interval $[0,1]$. $\tau_{S}^{2}$ and $\tau_{T}^{2}$ are overall spatial and temporal variation parameters, respectively. The values for these parameters define the strength of the simulated spatio-temporal effect. For $\rho_{S}$ and $\rho_{T}$ we consider low, medium and high  values, corresponding to $0.09$, $0.5$ and $0.9$, respectively. For $\tau_{S}^{2}$ and $\tau_{T}^{2}$, we also consider low, medium and high values, corresponding to $0.009$, $0.8$ and $3$ (large enough for this problem), respectively.  Overall, there are $3^{4} =243$ possible scenarios.
For presentation, we consider just a few  selected ones  for  $\tau_{S}$, $\tau_{T}$ and $\rho_{S}$,  $\rho_{T}$ as described in Table \ref{scenario_table_long} in section \ref{SM}.
For sensitivity analyse about the structure of the spatio-temporal random effect, we vary the values
of  $\tau_{S}$, $\tau_{T}$, $\rho_{S}$ and $\rho_{T}$ for each scenario. We also simulated different types of spatial models, all of
 which include spatial heterogeneity and spatial autocorrelation, whose strength change over time.
Additionally, we have also simulated several independent variables instead of just $3$ for sensitivity
analysis (data not shown). For each generated data, $3$ candidate regression models are fitted and
compared with respect to the comparison methods listed in subsection \ref{compare_section}: MLM CL3, MLM CAR ANOVA and MLM CONV.

For  cross-sectional analyses, we base our simulation on equation (\ref{two_level}).
We simulate two predictor
variables: one variable defined at individual level and the other defined at area level, both described
by  normal distributions. The individual level error term is also simulated from a normal distribution.
More details on the model equations used for simulating the spatial random effect are given in
section \ref{SM}, equation (\ref{simulate_outcome_cs}).
$\rho$ is an autocorrelation parameter and lies between $0$  and $1$. If $\rho$ is closer to $0$, then the simulated spatial effect is more similar to  spatial heterogeneity. If $\rho$ is closer to $1$, then this is more similar to a CAR structure. A value of $\rho$ between $0.4$ and $0.6$
corresponds to both medium spatial heterogeneity and autocorrelation.
The different $9$ scenarios are described in Table \ref{scenario_table_cs} in section \ref{SM}.
Here we just include $3$ scenarios for common data situations. For sensitivity analysis, for each scenario, different values of $\rho$ and $\tau^{2}$ are given.
For each generated data,  $3$ candidate regression models are fitted and compared with respect to the comparison methods listed in subsection \ref{compare_section}: MLM CL2, MLM CAR  and MLM RCAR.

\subsection{Computation}
All models are fitted in the Bayesian setting with Markov Chain Monte Carlo (MCMC) simulation
methods. All parameters whose full conditional distributions have a
closed form distribution, i. e. the regression parameters and all area level, individual and observational
level variance parameters, are updated using a Gibbs sampler (\cite{Gelfand}). The spatial and temporal parameter $\rho_{S}$ and $\rho_{T}$ for the MLM CAR ANOVA are updated using the slice sampler (\cite{Neal}).
Full conditional distributions for parameters of interest are available upon request. For a better comparability,   we relied on the expert system of BUGS (\cite{BUGS}), particularly run from within the R software using the  R-package R2WinBUGS (\cite{r2winbugs}).
We  used devices such as centering of the covariates as well as  hierarchical centering of the random effects (\cite{hier_centering})  to  reduce
correlation in the joint posterior and increase Markov Chain Monte Carlo
(MCMC) effective sample sizes.
To access  convergence and consistency of the chains,  single as well as two  parallel chains initialized at different points were used, and the  Geweke diagnostic (\cite{Geweke}), Brook $\&$ Gelman diagnostic (\cite{brook}) and  Heidelberger $\&$ Welch's  diagnostic (\cite{Heidelberger}) were applied.
 We ran the chains and chose the number of iteration until at least $\hat{R}$, the measure of mixing chains, is less than $1.02$ for all parameters and quantities of interest. For the final estimation, we used a single long run after discarding a part:
The length of the burn-in period was determined for each
model separately.

For the longitudinal analyses, the MLM CL3 (3 level) as well as the MLM CONV stabilized earlier at about $8000$ iterations. For the MLM CAR ANOVA, we needed up to $250000$ simulations for the
chains for some parameters to stabilize. We thinned the chain by storing only every $10$th draw.

For the cross-sectional analyses, we also thinned the chains by storing only every $10$th draw for
 the MLM CL2, MLM CAR and MLM RCAR in order to decrease autocorrelation and speed up
'mixing'. For the Markov chains for the MLM CAR and the MLM RCAR, we could not detect
departure from convergence after the $15000$th iterations. The MLM CL2 stabilized earlier.

 The models were run  in parallel using the  R-package  'batchtools' (\cite{batchpack}), which  provides a parallel implementation.
 The complete R-code for the simulation study is available upon request.

\section{Results for the  simulation studies}
\label{results}

\subsection{Results for longitudinal data}

In summary, MLM CONV model and the MLM CAR ANOVA model perform much better than the MLM CL3 model. This is particularly pronounced in case of a strong spatial variation
 and a changing spatial structure over time. Otherwise, the MLM tCARs should be used cautiously
to avoid overfitting, for instance.

\begin{center}
\includegraphics[width=15cm, height=14cm]{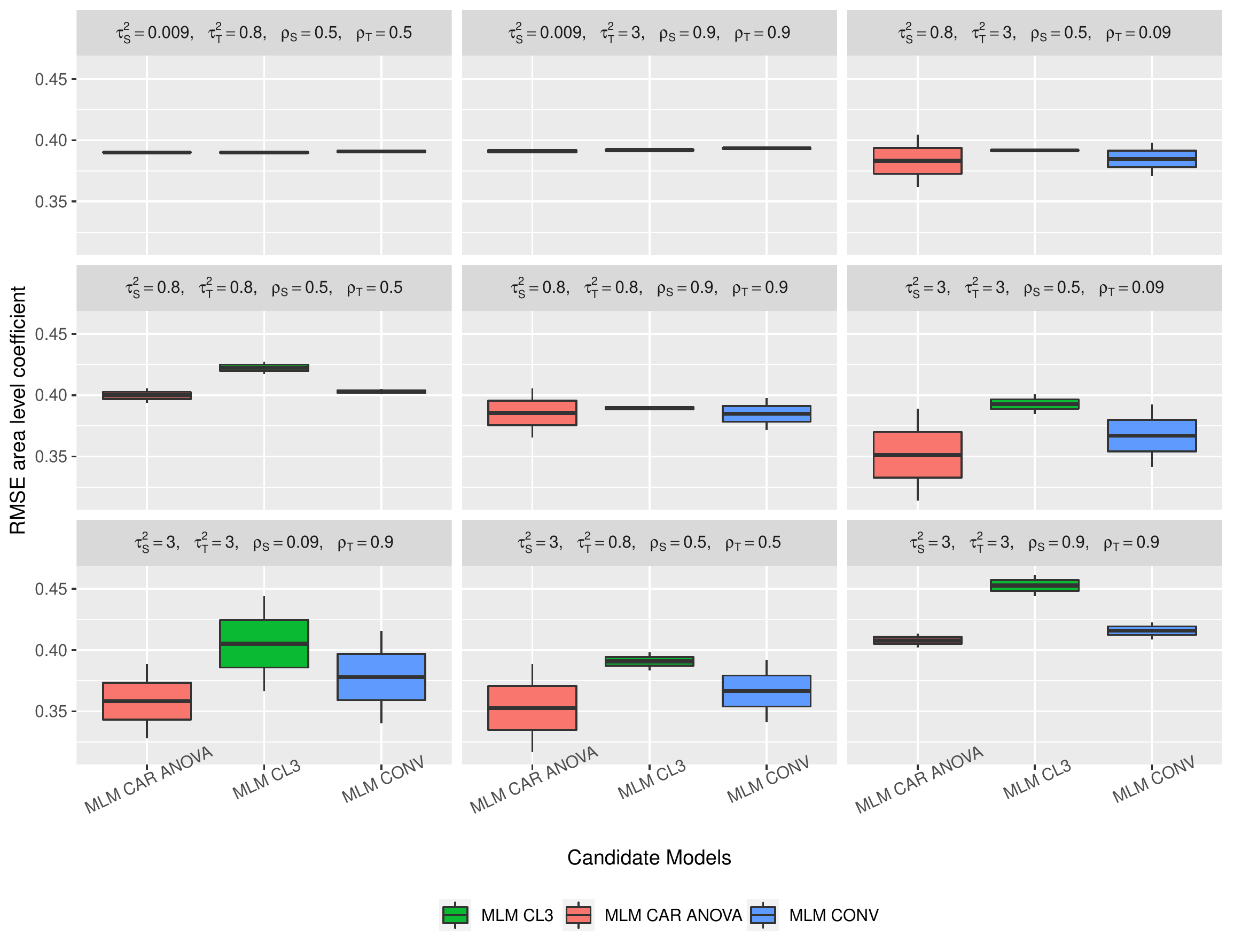}
\captionof{figure}{Comparison of the Root Mean Square Error (RMSE) for the area level variable coefficient,  for the set of selected scenarios of the simulated spatio-temporal effect. The true value  for the area level coefficient is $0.39$.  $\tau^{2}_{S}$ and $\rho_{S}$, $\tau^{2}_{T}$ and $\rho_{T}$ are overall  variance and autocorrelation parameters from equation (\ref{psi_long}) in section \ref{SM}, for space and time respectively.}
\label{RMSE_area_longitudinal}
\end{center}

In more details,
the individual level regression coefficients are very well retrieved by the $3$ candidate models in any of the  scenarios as indicated in Figure \ref{ind_coef}. However, the RMSE is more pronounced for the   MLM CL3 compared to the MLM tCARs  in general. The coverage of the $95\%$ Credible Interval (CI) is $100\%$ for all scenarios and for all candidate models (data not shown).
For the  area level variable coefficient in Figure \ref{RMSE_area_longitudinal}, the RMSE is not negligible for the three models. The bias and, therefore the RMSE is larger  in case of larger values of  $\tau^{2}_{S}$.  The RMSE is still larger for the  MLM CL3. The time coefficient is almost equally retrieved for the three methods, and the RMSE is larger  when the value of $\rho_{T}$  larger is (see Figure \ref{time_coef} in section \ref{SM}).

Observing Figures  \ref{DIC_fit}  and   \ref{loglikelihood_fit} (see  section \ref{SM}), the DIC and log-likelihood, respectively, for the MLM CAR ANOVA   and MLM CONV  are larger  than that of the  MLM CL3 in general. The DIC is smaller for the MLM  CONV compared to the MLM CAR ANOVA, and smaller for the MLM CAR ANOVA compared to the  MLM CL3. This indicates a better fit for the MLM tCARs in general.
Note from the results of the sensitivity analysis for the model parameters and model
goodness of fit that very small simulated observational level variances result in negative DIC values
(results not shown) even though fitting a correct model.

\begin{center}
\includegraphics[width=15cm, height=14cm]{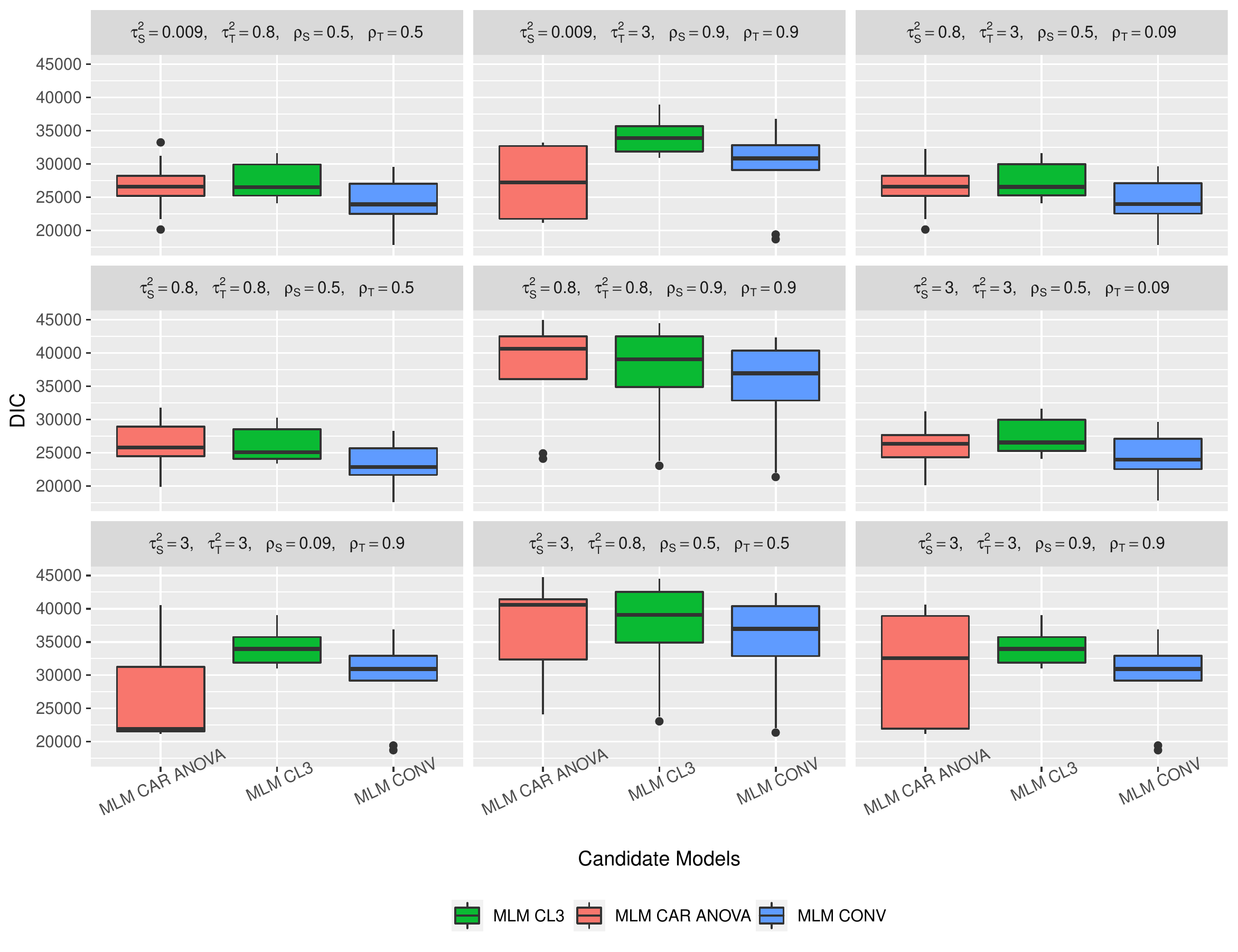}
\captionof{figure}{Comparison of the DIC, for a set of selected scenarios of the simulated spatio-temporal effect, longitudinal. $\tau^{2}_{S}$ and $\rho_{S}$, $\tau^{2}_{T}$ and $\rho_{T}$ are overall  variance and autocorrelation parameters from equation \ref{psi_long}, for space and time respectively.}
\label{DIC_fit}
\end{center}

In any case, the DIC values for the MLM tCARs are smaller in general. Moreover, small observational level (simulated) variances lead to very small bias in the individual level regression coefficients.  The coverage of the $95\%$ CI is  very good for the  three models in case of small values for $\tau^{2}_{S}$ and $\tau^{2}_{T}$ .

For very small values of spatial and temporal parameters, the MLM CL3 has smaller
 RMSE values for the regression coefficients in general compared to the MLM CAR ANOVA. This is
understandable as the MLM CAR ANOVA may be too complicated to fit such a simple structure.
The MLM CAR ANOVA in comparison to the MLM CONV models seems to be more complex. The
MLM CONV model shows  better results for the RMSE for coefficient estimates
and model fit in almost all scenarios.

For sensitivity analyses, different values for the fixed effect parameters for simulation were investigated. The results were similar regarding the behaviors of regression coefficients.

\subsection{Results for cross-sectional data}

The summary result is that the RMSE for the area level regression coefficients is larger for the
MLM CL2  and the MLM CAR, compared to the MLM RCAR. This depends on the strength of the
overall variance and spatial autocorrelation (CAR-structure) in the simulated data. The difference in
the RMSE is mainly due to the standard error, since there is little bias in the regression coefficients.
The individual level regression coefficients are not influenced very much. For a very small value of
the overall spatial variance parameter $\tau^{2}$, the bias is in general  negligible.
In more details, the
 individual level regression coefficients are very well retrieved by the three candidate models in any of the
scenarios, though the classical model performs worse; compare the RMSE for the individual level
variable in Figure \ref{RMSE_Int_area}. The coverage of the $95\%$ CI is $100\%$ for all scenarios and for all candidate
models (data not shown). For the area level variable coefficient in Figure \ref{RMSE_Int_area}, there is negligible bias
and RMSE for the three models in case of very small value of  $\tau^{2}$.  For moderate and higher values of  $\tau^{2}$,  the RMSE  is larger for the three models. Again, the RMSE is larger for the MLM CL2, smaller for the MLM RCAR and larger  the larger $\tau^{2}$ is.
The individual level error variance is better retrieved by the MLM CARs (data not shown).
The spatial autocorrelation parameter (data not shown) is also well retrieved by the MLM CAR and MLM RCAR; the  MLM CL2  tends to overestimate the corresponding random effects variances.

Observing Figure \ref{goodness_fit} (section \ref{SM}),
the DIC is smaller for the MLM RCAR compared to the   MLM CAR and MLM CL2.
The  log-likelihoods  are about the same for the three models.

 With additional explanatory variables (data not shown), we obtain similar results. For a stronger correlation between the area level variable and the CAR random effect term, the  MLM RCAR clearly has better performances.

\vfill
\vfill
\vfill
\vfill
\vfill
\vfill
\vfill
\vfill
\vfill

\begin{center}
\includegraphics[width=18cm, height=15cm]{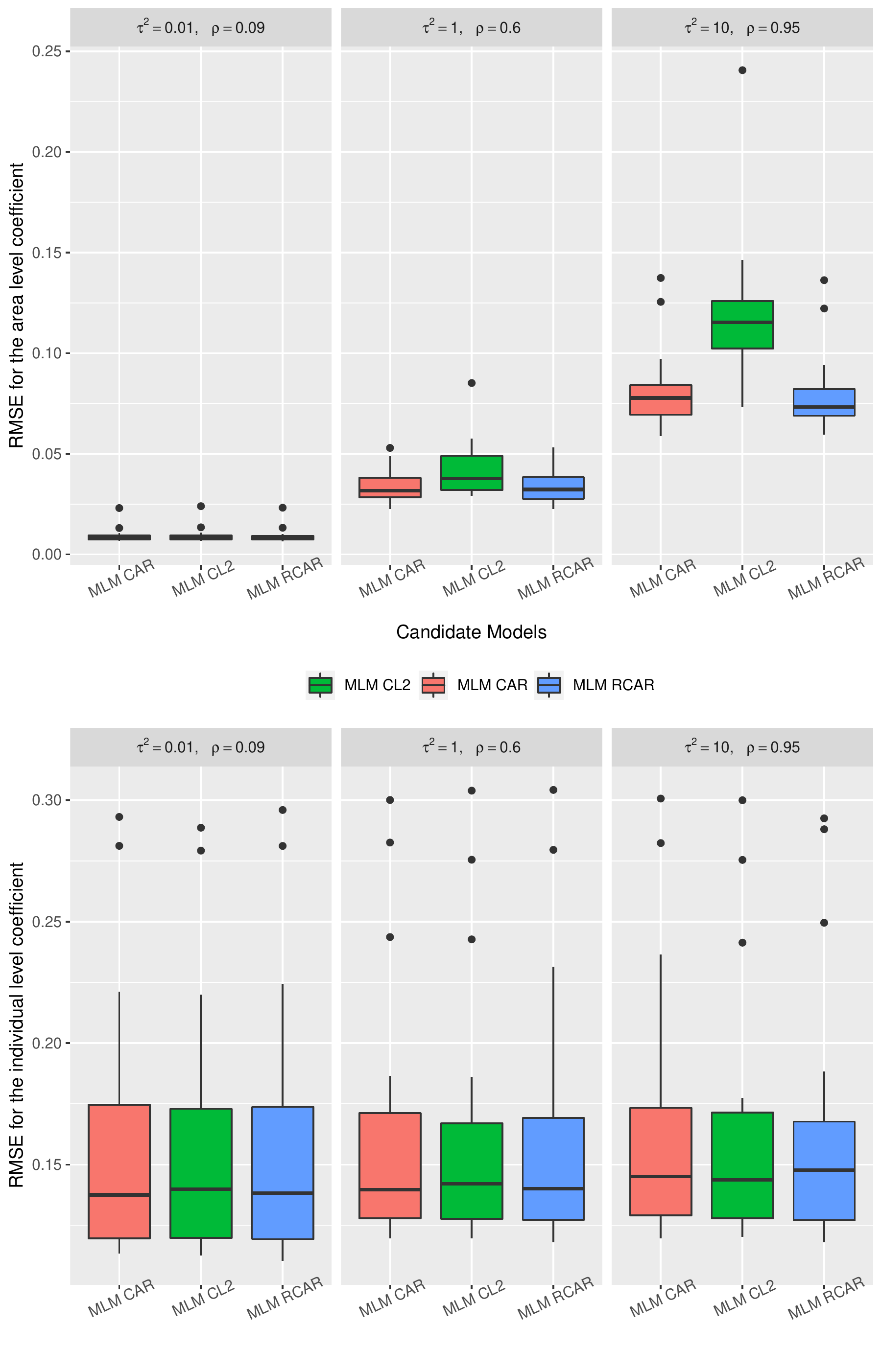}
\captionof{figure}{Comparison of the Root Mean Square Error (RMSE) for the area level and individual level variable coefficients, for a set of selected scenarios of the simulated spatial effect. The true value for the individual level coefficient is $-1.50$ while the true value for the area level coefficient is $0.14$. $\tau^{2}$ and $\rho$ are the overall spatial variance and autocorrelation parameters from equation (\ref{simulate_outcome_cs}) in section \ref{SM}, respectively.}
\label{RMSE_Int_area}
\end{center}

After   centering of the random effects, the computation time for the MLM RCAR  is slightly  larger than that of  the MLM CAR model. This, however, is not considerable compared to the risk of making weak inference with biased regression coefficients.

\newpage
\section{Decision tree}\label{guideline}

Although there exist many advanced methods available for analysing health data within
a geographical context, simpler methods are
often called upon to obtain a 'quick' solution. However, the choice of simplicity  sometimes may lead to erroneous conclusions. Based on the results of our simulation studies, we suggest the decision tree in Figure \ref{decision_tree} to help the users  analyse epidemiological data for which participants are nested within geographical  areas.
 We use a tradeoff between bias/RMSE, model fit and computational time to select the appropriate models. Overfitting in regression analysis can also produce misleading  regression coefficients and p-values (frequentist). We recommend  the MLM tCARs for longitudinal data instead of the MLM CL3 only when a strong spatial effect is expected in the data. Some of the MLM tCARs, like The MLM CONV  could be used routinely, no matter how strong the spatial effect is. For cross-sectional data, we recommend the use of the MLM RCAR and MLM CAR instead of the  MLM CL2 except when a very weak spatial effect is expected in the data.

For software implementation, the classical linear model, the MLM CL2, the MLM CL3 are available in software packages. The implementation for MLM tCARs and MLM CARs is not always given in software packages. See Table \ref{software_flow} in the supporting material in section \ref{SM} for detailed information on  the implementation of the methods described in this decision tree in software packages.

\vfill
\vfill
\vfill
\vfill
\vfill
\vfill
\vfill
\vfill
\vfill

\begin{center}
\includegraphics[width=35cm,
  height=19cm,
  keepaspectratio,]{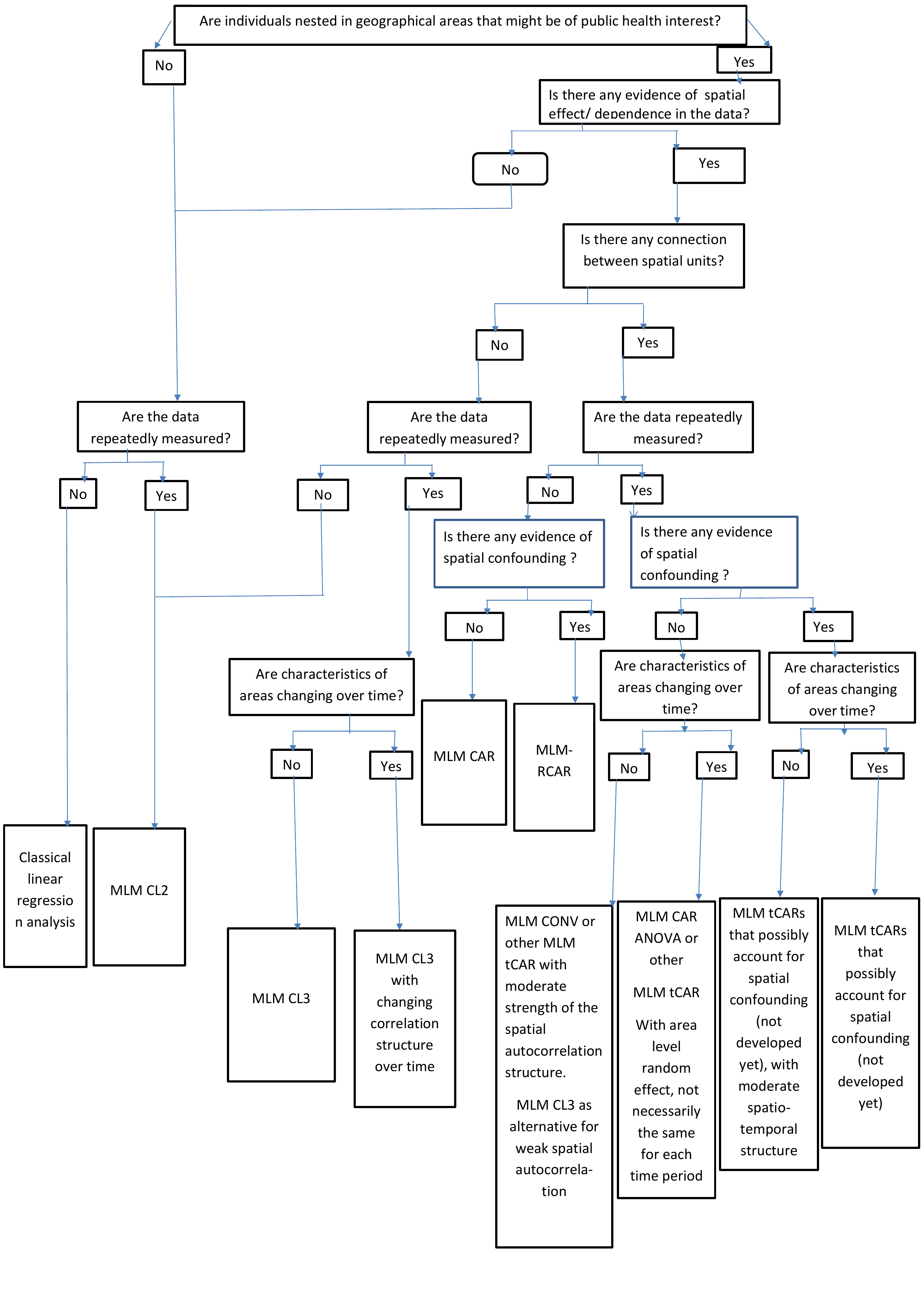}
\captionof{figure}{Decision tree for users, to analyse data for which participants are nested within geographical  areas that might be of public health interest. The method suggestion is based on a tradeoff between accuracy to retrieve regression coefficients, model's goodness of fit, and time needed to complete the fit.}
\label{decision_tree}
\end{center}

\section{Application to the analysis of the association between depression and greenness in the Heinz Nixdorf Recall Study}\label{application}

\subsection{Data}
The Heinz Nixdorf Recall Study (HNRS) is an ongoing population based cohort study from the cities  Essen, M{\"u}lheim and Bochum of the metropolitan Ruhr Area in Western Germany. The baseline was performed between $2000$ and $2003$ including $4814$ randomly selected men and women aged $45$ to
$75$ years old (\cite{Ester}).
Participants were invited to the study center three times each after 5 years ($t_{0}$, $2000-2003$; $t_{5}$, $2006-2008$; $t_{10}$, $2011-2015$). In addition, a yearly postal follow-up between the examinations was provided, which allows for more than 18 years of observational data.
The HNRS was approved by the
local ethics committees, and all participants
gave informed consent prior to participation.

Depressive symptoms during the previous week were assessed using the 15-item short-form questionnaire of the
Center for Epidemiologic Studies Depression
Scale (CES-D) (\cite{Hautzing}, \cite{Radloff}). For our analysis we used CES-D data of eight measurements, assessed between $2000$ and $2013$. Scores for
the 15-item version range from 0 to 45, with a
higher score indicating  more
frequent depressive symptoms.

Recent studies suggest evidence of the importance of green spaces for mental health. Green spaces  improve well-being or reduce physiological stress indicators (\cite{Ester}).
Exposure to green space is commonly measured either as surrounding greenness or access to green space. Here, we employ  the Normalized Difference Vegetation Index (NDVI) derived from satellite imagery (\cite{ndvi_scale}). Values of NDVI range from $-1$ to $+1$. Those around -1 generally correspond to water, while values near $0$ represent bare surfaces, e.g. rock, rooftops and roads or very scarce or dry vegetation.
Values approaching $1$ represent dense vegetation, e.g. rainforest.
 We base our analysis on eight time points of satellite imagery data.

Further covariates are included. Some  are directly measured at the district level like unemployment rate in district, obtained from the local census authorities of the respective cities of Bochum, Essen, and M\"ulheim.
 Unemployment is a strong indicator for material deprivation in a neighborhood and was used
  as an indicator of neighborhood-level socio-economic status (SES).
  Other covariates such as
 socioeconomic (e.g., income), demographic (e.g., age), gender, medical history and Body Mass Index (BMI) are measured at the individual level. In contrast to the NDVI, further covariates are time-invariant.

\subsection{Analysis and results}
We apply the MLM CL3, the MLM CAR ANOVA and the MLM CONV comparatively.
The  square root of the depression scores  leads to  models that better fit the assumptions and requirements of linear models. Square rooted depression is  continuous with values between
$0$ and $6.70$, has mean $2.47$, and a standard deviation of $1.20$.

Individual profiles of depression data with average trend line (Figure \ref{ind_prof_av_trend}, section \ref{SM}) are used exploratorily   to detect a slightly  linear  decreasing trend.
We also  analysed  the spatial confounding at baseline.
In order to decrease autocorrelation and
speed up 'mixing' of the MCMC, we thinned the chains by storing only every 10th
draw. The length of the burn-in period was determined for each
model separately.
For the Markov chains,  for the MLM CL3, we could not detect any departure from convergence
after the $8000$th iteration.
For the MLM CONV, the chain stabilized also very quickly, at about 9.000 iterations.
For the MLM CAR ANOVA, more iterations are needed, e.g., about $1000000$ to observe no departure from convergence for the spatial autocorrelation  parameter.

Figure \ref{post_plot_fixed} in the supporting materials shows the mixing of the chains for some fixed effect parameters of  the MLM CONV. The trace plots look similar for other methods.
 Trace plots, density plots and  further diagnostics for all parameter estimates as well as variance components are available upon request.

\begin{center}
 \begin{tabular}{|c|p{1cm}p{1cm}p{1cm}|p{1cm}p{1cm}p{1cm}|p{1.2cm}p{1.2cm}p{1.2cm}|}
 \hline
 &   \multicolumn{3}{c|}{\text{MLM}  \text{CL3}} &  \multicolumn{3}{c|}{\text{MLM}  \text{CAR ANOVA}}  &  \multicolumn{3}{c|}{\text{MLM}  \text{CONV}} \\ \hline %
      & $Med.$  & $2.5\%$  & $97.5\%$  &$Med.$& $2.5\%$ & $97.5\%$  &$Med.$& $2.5\%$ & $97.5\%$ \\\hline %
     intercept & $2.45$ & $2.40$& $2.49$ &       $2.75 $ &$0.85 $&$6.75 $           &$2.45$&$2.40$&$2.51$\\\hline
     greenness & $-0.32$ & $-0.47$ &$-0.16$      &$-0.32$&$-0.48$&$-0.16$ &     $-0.32$&$-0.47$&$-0.15$\\\hline
     female vs male  &$0.29$& $0.23$ &$0.34$     &$0.28$&$0.23$&$0.34$&       $0.28$&$0.23$&$0.34$\\\hline
     baseline Age &$0.009$& $0.005$& $0.01$ &      $0.009$&$0.005$& $0.01$ &        $0.009$&$0.006$&$0.01$\\\hline
     BMI &$0.02$& $0.01$& $0.02$ &        $0.02$&$0.01$&$0.02$  &     $0.02$&$0.01$&$0.02$\\\hline
      time & $-0.02$&$-0.02$ &$-0.01$&           $-0.02$&$-0.02$&$-0.01$ &     $-0.02$&$-0.02$&$-0.01$\\\hline
     unempl. in district &$0.03$&$0.02$ &$0.03$&          $0.03$&$0.02$&$0.03$        &$0.02$&$0.02$&$0.03$\\\hline

    $\tau_{S}$ & $-$ &$-$ &$-$         &$0.07$&$0.04$&$0.13$         &$ $&$ $&$ $\\\hline
     $\tau_{T}$ & $-$  & $-$ & -        &$0.07$ &$0.04$&$0.15$       &$ $&$ $&$ $\\\hline
    $\rho_{S}$ & $-$  &$-$ & $-$        &$0.68$&$0.13$&$0.99$      &$ $&$ $&$ $\\\hline
      $\rho_{T}$& $-$ & $-$ & $-$        &$0.82$&$0.03$&$0.99$         &$ $&$ $&$ $ \\\hline
    $\sigma_{r_{0}}$  &$0.85$& $0.83$& $0.88$      &$0.85$&$0.83$&$0.88$        &$0.85$&$0.83$&$0.87$\\\hline
    $\sigma_{r_{1}}$  &$0.09$& $0.09$& $0.10$         &$0.09$&$0.09$&$0.10$        &$0.09$&$0.09$&$0.10$\\\hline
    $\sigma_{\omega}$   &$-$& $-$& $-$      &$0.06$&$0.04$&$0.09$          &$-$&$-$&$-$\\\hline
   $\sigma_{u_{0}}$   & $0.06$ & $0.04$ & $0.09$       &$-$&$-$&$-$          &$-$&$-$&$-$\\\hline
      $\sigma_{e}$ & $0.71$ & $0.70$ & $0.72$       &$0.71$&$0.70$&$0.72$           &$0.71$&$0.70$&$0.72$\\\hline
      median $\sigma_{\omega t}$ & $-$ & $-$ &$ - $    &$-$&$-$&$-$       &$0.09$&$0.05$&$0.16$ \\\hline
      median $\tau_{t}$ & $-$ & $-$ &$ - $    &$-$&$-$&$-$     &$0.09$&$0.05$&$0.22$ \\\hline
       log-likelihood &  & $-27194.85$ &     &&$-27190.52$&         &&$-27186.08$&\\\hline
      \text{DIC} &  & $59570$ &     &&$59570$&         &&$59560$&\\ \hline
 \end{tabular}

\captionof{table}{Posterior quantiles from equation (\ref{MLMCAR}) for the MLM CL3  , MLM CAR ANOVA and MLM CONV  for the analysis of the association between greenness and depressive symptoms of the Heinz Nixdorf Recall Study.}
\label{result_application}
\end{center}

The results suggest a negative association between greenness and depressive symptoms for all three methods ($0$ is not included in the CI) as indicated in Table \ref{result_application}.  For MLM CL3, MLM CONV  and MLM CAR ANOVA respectively, a unit increase in the value of NDVI leads to a decrease of the root of depression score, on average by  $-0.32$ ($[ -0.47; -0.16]$, $[ -0.48; -0.16]$, $[-0.47; -0.15]$). The time slope is on average $-0.02$ ($[-0.03;-0.02]$) for all three methods ($0$ not included in the CI). This indicates  a slightly linear decreasing  trend, which is in perfect accordance with the
 exploratory individual profile plot of the data in Figure \ref{ind_prof_av_trend} of section \ref{SM}.
We also notice that the coefficient estimates for  the area level variables are  the same for the three models.
The intercept is slightly different for the MLM CAR ANOVA.
 This is also in line with the results of the simulation study, where in case of a very small overall spatial effect, not changing much over time, and a medium spatial autocorrelation, the three compared methods give approximately the same results for coefficient estimates and credible intervals.
  In fact, the spatial heterogeneity parameter  for the MLM CONV  $\sigma_{\omega_{t}}$ is very small  (median value $0.09$), and the spatial autocorrelation parameter for the MLM CONV $\sigma_{\varphi_{t}}$  as well  (median value $0.09$). Both are  not changing much over time. This is also in accordance with the MLM CAR ANOVA, where the overall spatial and temporal parameters $\tau_{S}$ and $\tau_{T}$ are  $0.07$ and $0.06$.  A medium value of the spatial autocorrelation parameter $\rho_{S}= 0.68$ is indicative of a medium  spatial dependence. A large value of the  temporal autocorrelation parameter $\rho_{T}= 0.82$  is indicative of a spatial effect that is not changing much over time. The spatio-temporal interaction parameter $\tau_{\gamma} = 0.06$ is also very weak. Both the posterior log-likelihood and DIC  are better for the MLM CAR ANOVA and MLM CONV, particularly the MLM CONV, compared to the MLM CL3. The total $30000$ iterations took $57.26$ minutes for the  MLM CL3,  $57.22$ minutes for the MLM CONV  and $57.98$ minutes for the MLM CAR ANOVA (just for comparison)  on an ordinary personal computer.

  The similar results for fixed effect estimates were also expected from the decision tree after having a look at the spatial structure of the data. For a weak spatial effect, not changing much over time, the MLM tCARs were suggested, with the MLM CL3 as alternative.

\section{Discussion}
\label{discussion}
Our aim was to explore the effect of considering, or neglecting spatio-temporal random effects, on regression coefficients in epidemiological data, with participant  nested within geographical units.

To achieve this, we considered the classical multilevel models and  the multilevel conditional autoregressive models.
The multilevel conditional autoregressive models are obtained  by combining multilevel models and the properties of Markov Random Fields in order to borrow strength in adjacent areas.
We considered separately regression models for  cross-sectional studies as well as for longitudinal studies in simulation studies in several scenarios of the spatial and spatio-temporal  effects.
For cross-sectional analyses, we compared the MLM CL2 , the MLM CAR  and the MLM RCAR. For longitudinal studies we introduced new models named MLM tCARs (MLM CAR ANOVA and MLM CONV).  Such models are not common in the literature, at least not for epidemiological analyses. These models help to explicitly borrow strength and simultaneously account for individual dynamics as well as area dynamics. After introducing the MLM tCARs, we  compared them to the MLM CL3. In summary, the results indicated  that neglecting  either  the spatial or spatio-temporal effects  leads to larger  RMSEs in coefficient estimates  and worse model fit in general.
Furthermore, we provided a decision tree for model selection in
epidemiological studies for spatially nested data, based on simulation studies. At the end, we applied the methods to the analysis of the association between depression and greenness in the HNRS.
In the following, we discuss our results, pointing out
the strengths and limitations.

To achieve our goals, we had to simulate the spatial and spatio-temporal random effects. The simulation of random effects from CAR models is not common and should be taken cautiously
 (\cite{Old_new}). However, CAR type random effects are successfully used in many works and applications  to simulate the spatial structure in order to show the correctness and advantages of some CAR-like spatial models (\cite{CarBayes}).
We used the geography of the HNRS, but  the results can be generalized to other geographical structures.

For the application, the three models showed the negative association between greenness and depressive symptoms as well as  a  linear decreasing individual trend. This analysis of the association  at the individual level  is also perfectly  in line with the analysis by \cite{djeudeu}, which was performed on the HNRS at an aggregated level. Moreover,  the current  analysis  has the added value that all  variables are used at their finest  spatial level. This avoids the risk of the ecological and the atomic fallacy by aggregating or disaggregating them. Though the analysis of the association and growth showed the same result for all three models for the coefficient estimates of interest,  the MLM CAR ANOVA  as well as the MLM CONV   showed an overall better fit compared to the MLM CL3. Moreover,
using  the MLM CAR ANOVA  or the MLM CONV, we were able   to separate out the individual  time trend and the spatial time trend. The spatial effect was not changing very much over time. The overall spatial effect was medium and the spatial autocorrelation was rather strong. This explains why, the models showed almost the same behaviours for coefficient estimates as expected from the scenarios of the simulation study and as indicated in the decision tree in Figure \ref{decision_tree}.

The MLM CARs and MLM tCARs  have the limitation that using   them unnecessarily for  non-complex  data structures may lead to overfitting and slightly  time-consuming fits. However, the damage of neglecting the spatial structure is more important than the unnecessary computational burden to fit the model.
The bias for individual level coefficients are generally negligible for all methods. However, not accounting for the spatio-temporal dependence and dynamic could lead to large standard deviations for regression coefficients and therefore larger RMSEs for classical models.  The estimates of the standard errors determine the 'significance' of the fixed effect parameters (frequentist). Not choosing the right model to account for the spatio-temporal effect  will lead to (falsely) small p-values (frequentist) and, therefore, false 'significant' associations between health outcome and exposure in some epidemiological studies.
Therefore, it is recommended to consider a  model that accounts for the spatial effect as well as the dynamic of the
residual variation rather than  using the  MLM CL2 or MLM CL3.

The spatio-temporal CAR-prior should, however, be used cautiously.
In the situation of spatial confounding, we recommend  to use the restricted CAR models. This will avoid  the fixed and random area-level effects to compete to explain common variation, which could distort estimates of the fixed effects  area level  regression coefficients and unduly inflate their posterior variances. This was considered for our cross-sectional analyses.
The MLM CAR and MLM RCAR were just a bit more   time consuming compared to the MLM CL2. Although  this seems to be a huge disadvantage, in absolute practical terms in a multitasking computing environment, it makes little impact. Relative to the time taken to collect the data (sometimes more than $10$ years) it is irrelevant. Restricted MLM tCAR models still need to be fully developed and were not part of our simulation study.

Using MLM CARs and MLM tCARs, spurious inferences regarding fixed  effects parameters can be avoided. This is particularly important when the primary inferential focus is on fixed effect estimates as in several epidemiological analyses. If the goal of the analysis was model comparison in terms of prediction, we would have to use some sort of hold-out of data. See \cite{white_smooth} for analysing spatial data with the goal of spatial prediction. Methods suited for inference may not always be appropriate for comparing models in terms of  prediction (\cite{bias_reg_compare}).

We noticed that the MLM CONV  has a less complex spatio-temporal structure and performs a bit better than the MLM CAR ANOVA in most of the scenarios, also in the application. MLM tCARs with different spatio-temporal CAR prior structures  for the area level random effects can also be compared in  future works to further improve  the decision tree.
The MLM CAR ANOVA  did not retrieve the  intercept properly for certain scenarios. This could be a  result from    spatial and temporal confounding that should jointly be addressed in future studies.
The overall results in this paper are developed for Gaussian likelihood models and could easily be extended to other likelihood models (generalized linear models). A preliminary simulation study   with the logit-link function for cross-sectional data showed similar results for coefficient estimates that we described here.

\bibliographystyle{apalike}
\bibliography{Multilevel_CAR_Arxiv}

 \section{Supplementary material}\label{SM}

\subsection{Tables and figures}
\vspace{0.5cm}

\begin{tabular}{|p{1.5cm}|p{1cm}|p{1cm}|p{1cm}|p{1cm}|p{8cm}|}
  \hline
 Scenario No&  $\tau_{S}^{2}$  & $\rho_{S}$ & $\tau_{T}^{2}$ & $\rho_{T}$&  Meaning/Interpretation\\\hline
  $1$&$0.09$ & $0.5$ &0.8  & 0.5&weak spatial effect, medium spatial heterogeneity and autocorrelation, medium temporal effect \\\hline
  $2$&$0.009$ & $0.9$ & 3 & 0.9 &weak spatial effect,  mainly spatial autocorrelation, strong temporal effect\\\hline
  $3$&0.8 & $0.5$ & 3  & 0.09 &medium spatial effect,  medium spatial heterogeneity and medium spatial autocorrelation, strong temporal effect  \\\hline
  $4$&$0.8$ &$0.5$ & 0.8 &0.5 &medium spatial effect, medium spatial heterogeneity, medium temporal effect  \\\hline

  $5$&$0.8$& $0.9$ &$0.8$&$0.9$&moderate spatial effect, mainly spatial autocorrelation, medium temporal effect \\\hline
  $6$&$3$ & $0.5$ &$3$&$0.09$&strong spatial effect,  medium spatial autocorrelation, strong temporal effect\\\hline
  $7$&$3$& $0.09$ &$3$&0.9 &strong spatial effect,  mainly spatial autocorrelation, strong temporal effect  \\\hline
  $8$&$3$ &$0.5$ & $0.8$&0.5 &strong spatial effect,  medium spatial autocorrelation, medium temporal effect  \\\hline

  $9$&$3$ & $0.9$ &$3$&0.9& strong spatial effect, mainly spatial medium spatial autocorrelation, strong temporal effect \\\hline
\end{tabular}\vskip18pt
\captionof{table}{The selected scenarios of the simulated spatio-temporal effect.  Here, all the selected scenarios  are  presented in the paper.}
\label{scenario_table_long}

\begin{tabular}{|p{1.5cm}|p{2cm}|p{1.5cm}|p{8cm}|}
  \hline
 Scenario number& Value of $\tau^{2}$  & Value of $\rho$ &  Meaning/Interpretation\\\hline
  $1$& $1$ & $0.95$ & moderate spatial effect, mainly spatial autocorrelation \\\hline
  $2$&1 & $0.09$ &  moderate spatial effect, mainly spatial heterogeneity \\ \hline
 \cellcolor[gray]{.8}{$3$}&$1$&$0.6$&moderate spatial effect, medium spatial heterogeneity and medium spatial autocorrelation \\ \hline

  $4$&$10$ & $0.09$ &  strong spatial effect,  mainly spatial spatial heterogeneity \\\hline
   {\cellcolor[gray]{.7}}$5$&$10$ & $0.95$ & strong spatial effect,  mainly spatial autocorrelation\\\hline
  $6$&10 & $0.6$ &  strong spatial effect,  medium spatial heterogeneity and medium spatial autocorrelation  \\\hline
  $7$& $0.01$& $0.95$ &  weak spatial effect,  mainly spatial autocorrelation\\\hline

{\cellcolor[gray]{.7}}  8& $0.01$  &  $0.09$  &   weak spatial effect,  mainly spatial heterogeneity  \\\hline
   $9$&$0.01$ & $0.6$ &  weak spatial effect,  medium spatial heterogeneity and medium spatial autocorrelation \\
  \hline
\end{tabular}\vskip18pt
\captionof{table}{The different scenarios of the simulated spatial effect. The scenarios with gray color are the one presented in the paper.}
\label{scenario_table_cs}

\begin{table}
\resizebox{\columnwidth}{!}{%
\begin{tabularx}{24.45cm}{|p{3cm}|p{6cm}|p{7cm}|p{6.7cm}|}
\hline
 & \textbf{Statistical Model} & \textbf{Usual R packages, implementation in R}&
 \textbf{Usual implementation in SAS} \\ \hline
Cross-sectional design, classical non spatial methods&\begin{minipage}[t]{\linewidth}%
Classical multivariate linear, Poisson, Binomial, negative Binomial, log normal regression.\\
There are non-parametric and semi-parametric methods as well.
\end{minipage}  &  \begin{minipage}[t]{\linewidth}%
Frequentist approach include  R-packages stats (lm function for linear and glm for generalized linear model), MASS (glm.nb),  gamm4 (gamm, gamm4),  nlme (lme) lme4 (lmer, lmer2, glmer), glmmADMB (glmmadmb) for mixed effect, gee (gee) for marginal models.\\
Bayesian approach includes R package
MCMCglmm, R2WinBUGS, RINLA, rstan.\\
The list is not exhaustive.\\
\end{minipage}&
\begin{minipage}[t]{\linewidth}%
Frequentist approach includes  proc reg, proc glm for linear regression, and \\
proc logistic, proc genmod, proc glimmix, proc nlmixed for generalised linear models.\\
Bayesian approach include proc mcmc and  built-in capabilities in the genmod procedure. \\
\end{minipage}\\ \hline
Cross-sectional design, classical spatial methods&  \begin{minipage}[t]{\linewidth}%
Classical MLM (MLM CL2)\\
Fixed Effect Methods (FEM) \\
Genaralized Estimating Equations (GEEs)\\
\end{minipage}& \begin{minipage}[t]{\linewidth}%
Frequentist approach include gamm4 (gamm, gamm4)  lme4 (function lme),\\
mgcv (gam, semi-parametric).\\
Bayesian approach includes R packages
MCMCglmm, R2WinBUGS, RINLA, rstan.

\end{minipage}&
\begin{minipage}[t]{\linewidth}%
Frequentist approach include   proc mixed for linear regression, proc logistic, proc genmod and proc GEE for fixed or marginal effect, proc glimmix and proc nlmixed  for generalized ME,\\
Proc GAM for semi-parametric models\\
Bayesian approach includes proc MCMC. \\
\end{minipage}\\\hline
Cross-sectional design, spatial heterogeneity and spatial autocorrelation & \begin{minipage}[t]{\linewidth}%
Multilevel CAR (MLM CAR),\\
Multilevel resricted CAR (MLM RCAR),\\
Semi-parametric regression (generalized additive models).\\
\end{minipage} & \begin{minipage}[t]{\linewidth}%
Bayesian approach include R package
MCMCglmm, R2WinBUGS, RINLA, rstan, CARBayes, HSAR.
\end{minipage}&
\begin{minipage}[t]{\linewidth}%
Bayesian approach includes proc mcmc \\
\end{minipage}\\ \hline
Longitudinal design, classical methods & \begin{minipage}[t]{\linewidth}%
 Univariate and multivariate ANOVA,\\
Growth curve models, \\
 multilevel regression models (MLM CL3),\\
SEM (Structural Equation Modeling) for longitudinal data,\\
Fixed Effect Models (FEM),\\
Generalized Estimating Equations (GEEs),\\
Semi-parametric models.

\end{minipage} & \begin{minipage}[t]{\linewidth}%
Frequentist approach include   R packages  gamm4 (gamm, gamm4)  lme4 (lme)\\
geepack and gee for Generalized Estimating Equation, multgee for multinomial response, CRTgeeDR. \\
Semi-parametric methods includes mgcv (gam).\\
Bayesian approach includes R package
MCMCglmm, R2WinBUGS, RINLA, rstan.

\end{minipage}&
\begin{minipage}[t]{\linewidth}%
Frequentist approach include   proc mixed for linear regression, proc logistic, proc genmod and proc gee for fixed or marginal effect, proc glimmix and proc nlmixed   for generalized MLM.\\
Bayesian approach includes proc mcmc \end{minipage} \\ \hline
Longitudinal design, changing spatial effect, spatial autocorrelation and spatial heterogeneity& MLM tCARs for  longitudnal data&
 \begin{minipage}[t]{\linewidth}%
Implementation using WinBUGS run from within R via the R-package R2WinBUGS
\end{minipage}
 &\\ \hline
\end{tabularx}
}
 \caption{Summary of methods (and software packages) for the analysis of spatial data in cross-sectional and longitudinal design  in Epidemiology.}
  \label{software_flow}
\end{table}

\newpage
\begin{center}
\includegraphics[width=11cm, height=9cm]{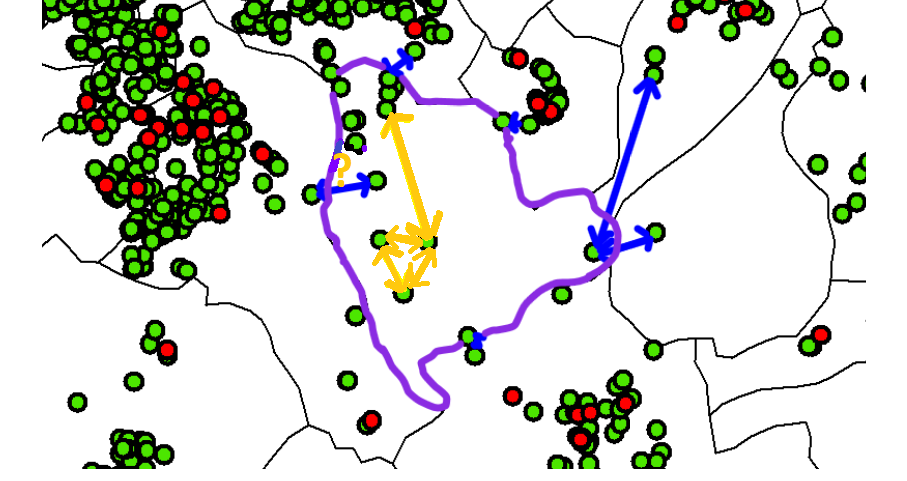}
\captionof{figure}{Description of  spatial effects for the example of the Heinz Nixdorf Recall Study:  The red/green points represent participants' positions with high/no high value of depression score, the yellow arrows indicate the
spatial proximity for participants in the same geographical unit, while the blue ones are for participants in adjacent units. Two participants in adjacent units may be closer and have  more similar outcome than participants in the same spatial
unit (yellow arrows). This implies that both spatial heterogeneity and spatial autocorrelation should be accounted for.}
\label{adjacency}
\end{center}

\begin{center}
\includegraphics[width=16cm, height=14cm]{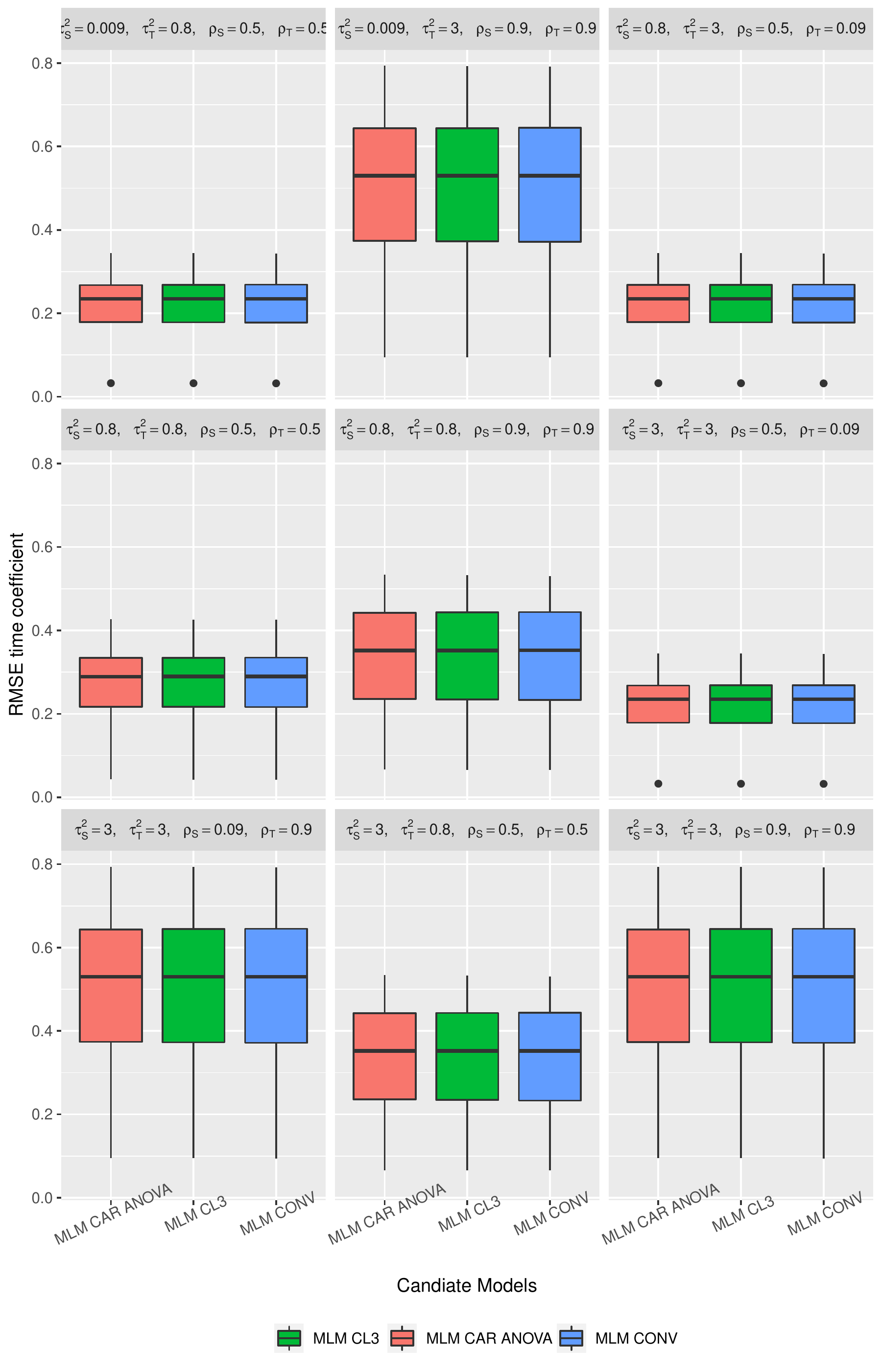}
\captionof{figure}{Comparison of the Root Mean Square Error (RMSE) for the time variable coefficient (growth),  for the set of selected scenarios of the simulated spatial effect. The true value  for the time coefficient is $-0.1$.  $\tau^{2}_{S}$ and $\rho_{S}$, $\tau^{2}_{T}$ and $\rho_{T}$ are overall  variance and autocorrelation parameters from equation (\ref{psi_long}), for space and time respectively.}
\label{time_coef}
\end{center}

\begin{center}
\includegraphics[width=16cm, height=14cm]{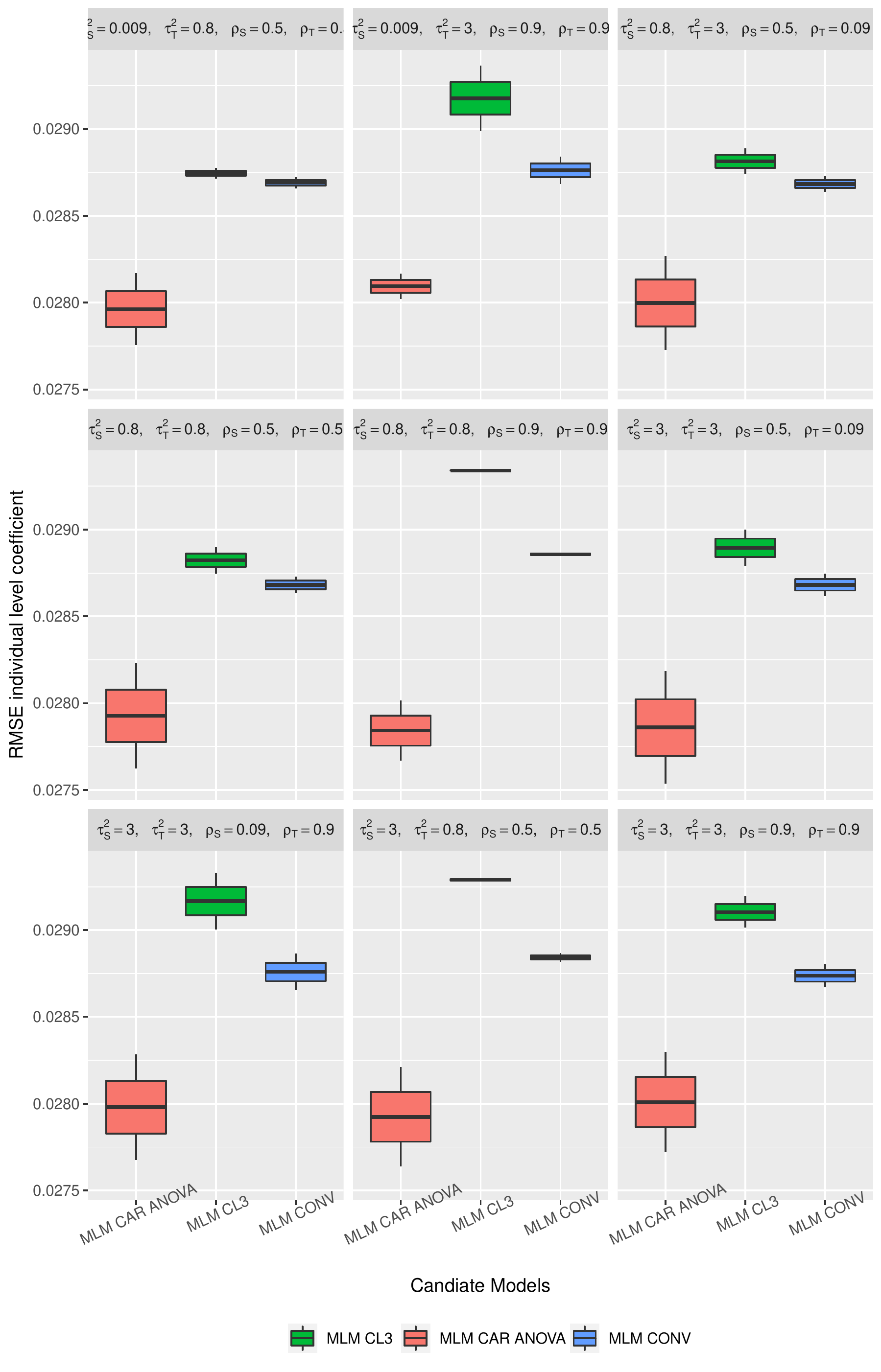}
\captionof{figure}{Comparison of the Root Mean Square Error (RMSE) for the individual level variable coefficient,  for the set of selected scenarios of the simulated spatial effect. The true value  for the individual level  coefficient is $-1.72$.  $\tau^{2}_{S}$ and $\rho_{S}$, $\tau^{2}_{T}$ and $\rho_{T}$ are overall  variance and autocorrelation parameters from equation (\ref{psi_long}), for space and time respectively.}
\label{ind_coef}
\end{center}

\begin{center}
\includegraphics[width=16cm, height=14cm]{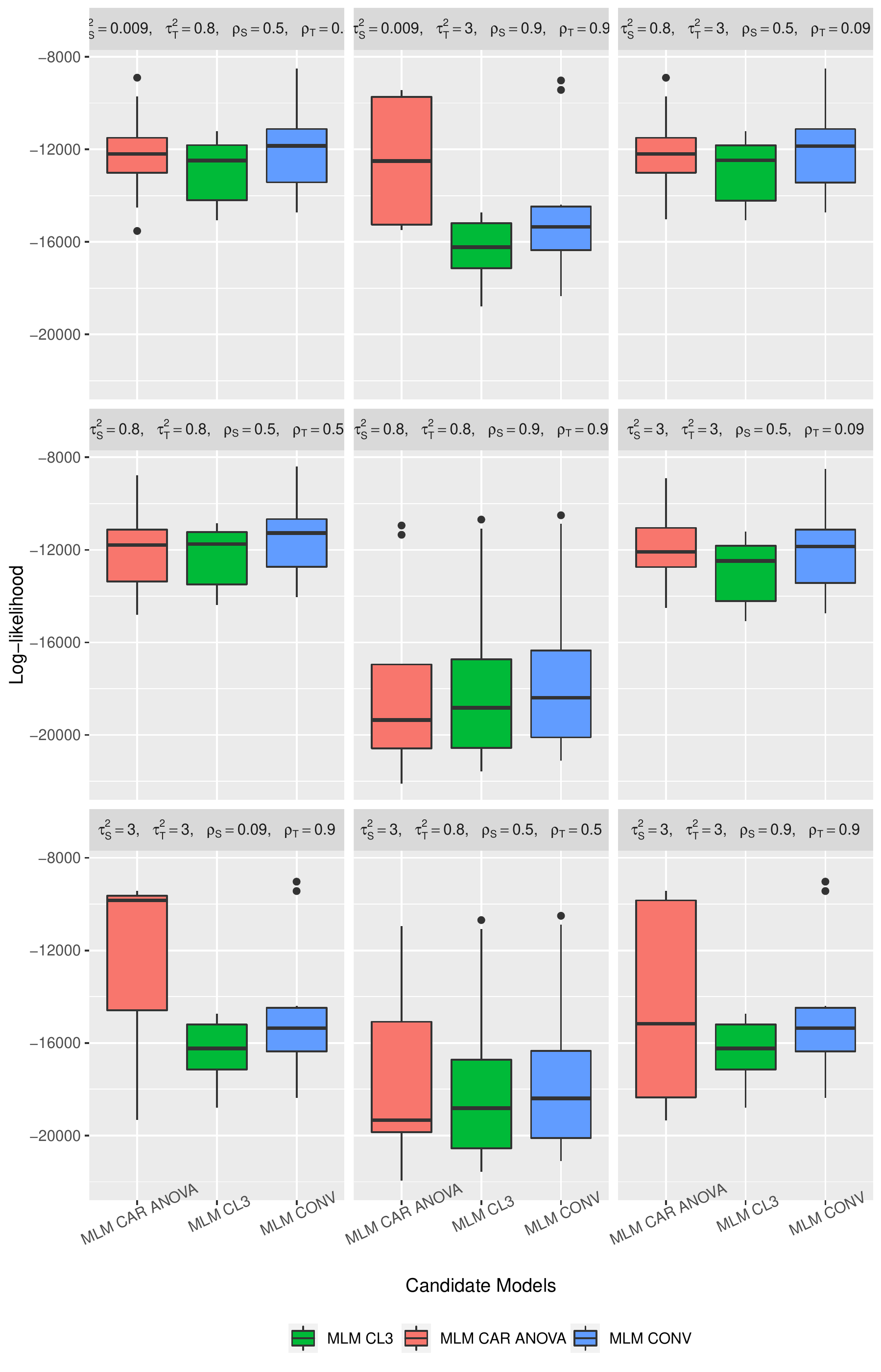}
\captionof{figure}{Comparison of the posterior log-likelihoods, for a set of selected scenarios of the simulated spatio-temporal effect for  longitudinal studies. $\tau^{2}_{S}$ and $\rho_{S}$, $\tau^{2}_{T}$ and $\rho_{T}$ are overall  variance and autocorrelation parameters from equation (\ref{psi_long}), for space and time respectively.}
\label{loglikelihood_fit}
\end{center}

\begin{center}
\includegraphics[width=16cm, height=14cm]{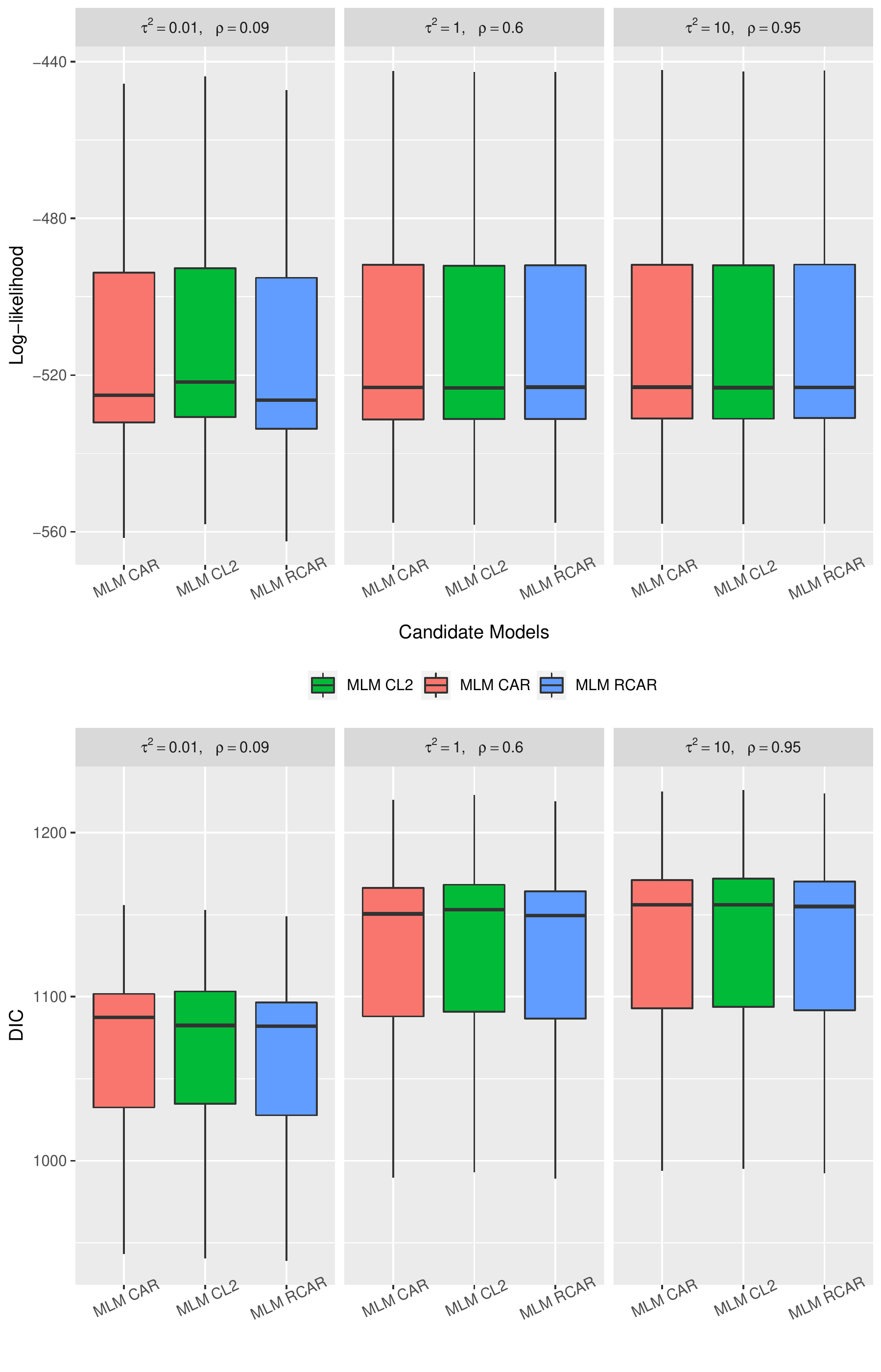}
\captionof{figure}{Comparison of the goodness of fit, for a set of selected scenarios of the simulated spatial effect for  cross-sectional studies. $\tau^{2}$ and $\rho$ are the overall spatial variance and autocorrelation parameters from equation (\ref{simulate_outcome_cs}), respectively.}
\label{goodness_fit}
\end{center}

\begin{center}
\includegraphics[width=12cm, height=10cm]{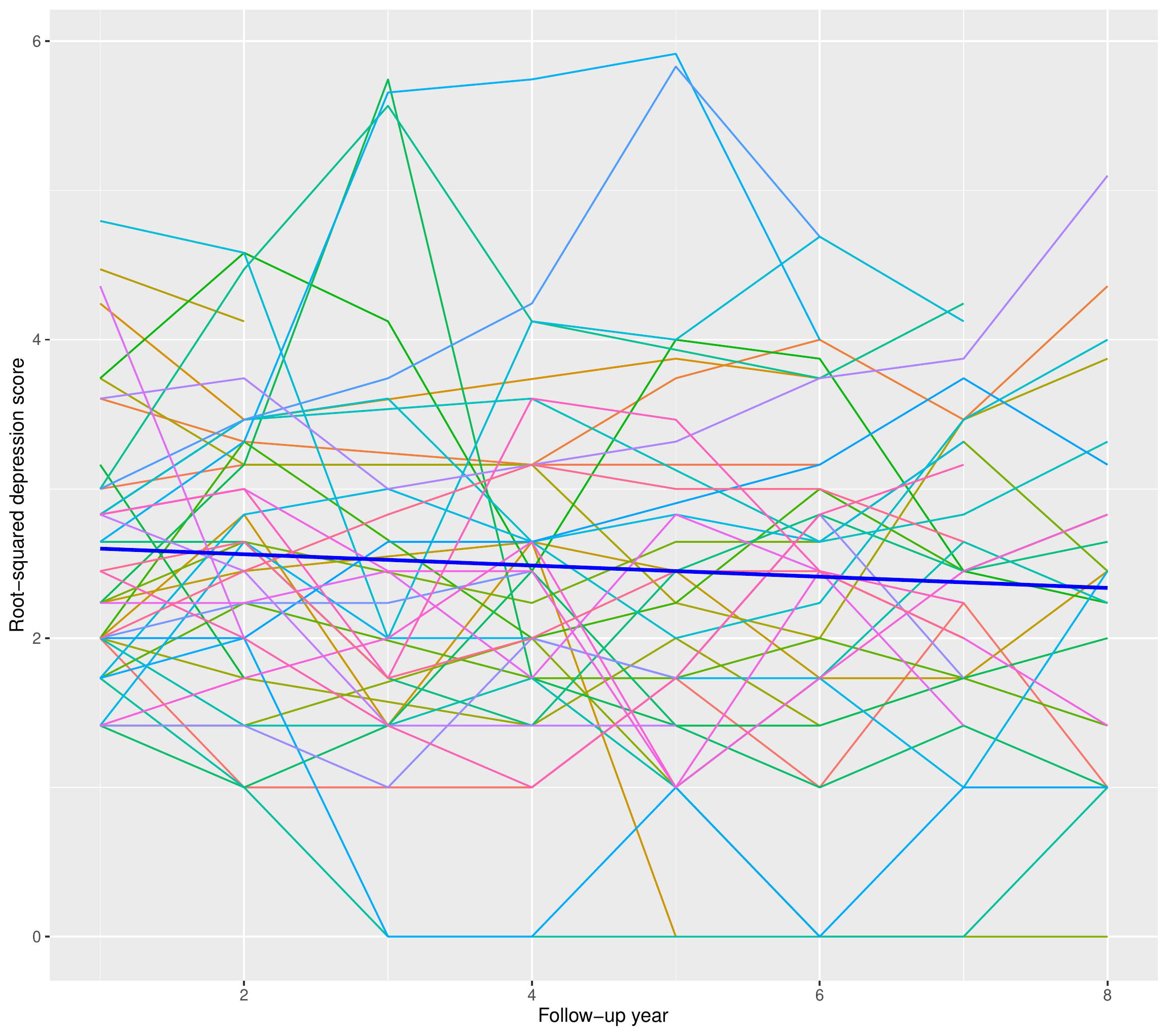}
\captionof{figure}{Exploratory plot  of the individual trajectories and mean trajectory for randomly selected 50 participants, to identify  average individual  trends within subjects. The  dense linear blue  line represents the means trajectory. It is indicative of a linear decreasing trend.}
\label{ind_prof_av_trend}
\end{center}

\begin{center}
\includegraphics[width=15cm, height=15cm]{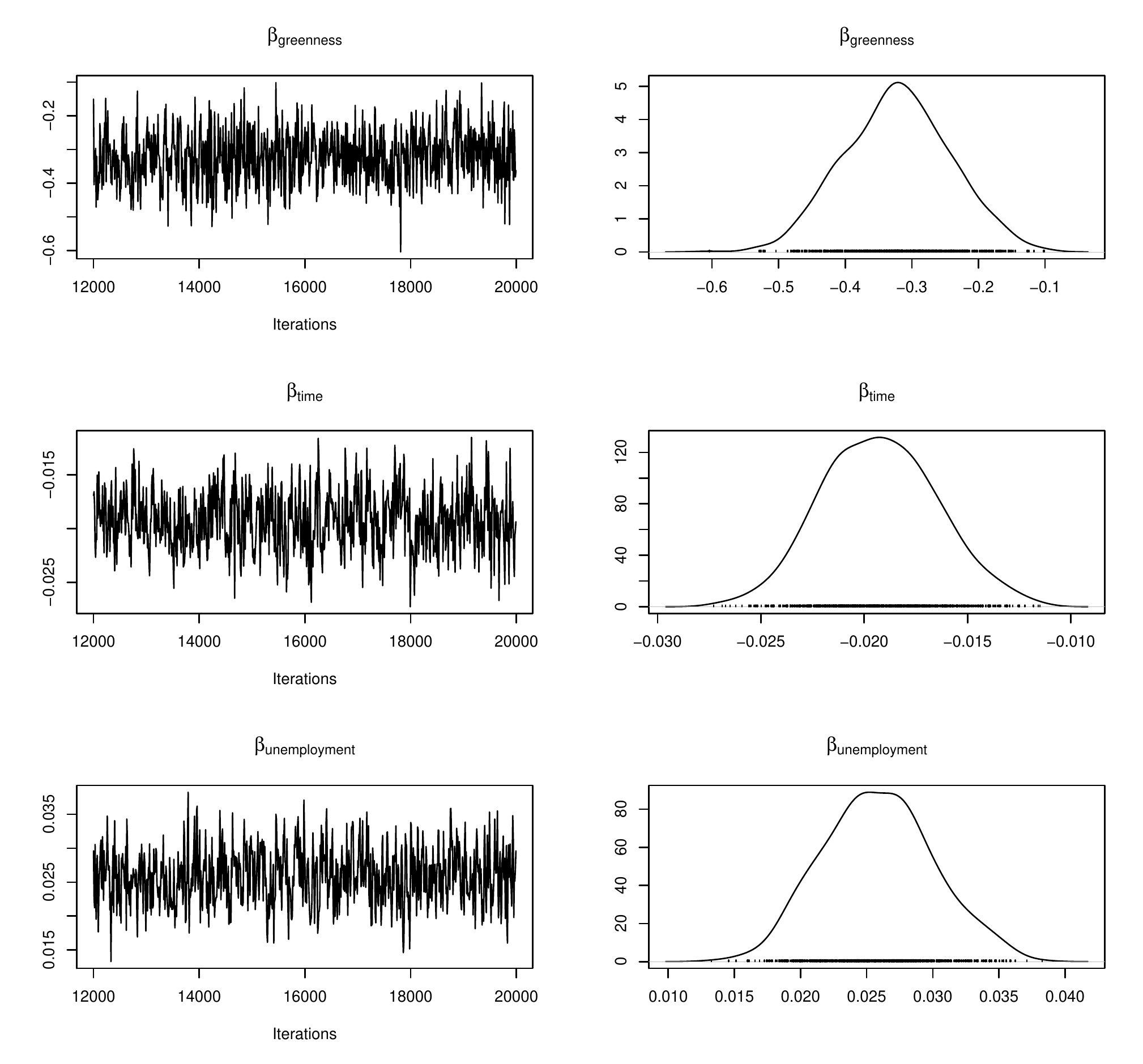}
\captionof{figure}{Post diagnostic plots (trace and density plots) of the fixed effect parameters of interest.}
\label{post_plot_fixed}
\end{center}

\subsection{Simulation of the spatio-temporal effect}
We use the following equation, step by step, to generate the random effect and then the dependent variable:

\begin{equation}\label{psi_long}
\left\{
  \begin{array}{ll}
   \Phi_{vec} &= (\Phi_{1}, \Phi_{2},\ldots, \Phi_{N})^{T} ~~\text{with} ~~ \Phi_{t} \sim   \mathcal{N}_{K}(0_{K}, \tau^{2}_{S} Q^{-1}), ~~ t=1, \ldots,N,\\
     \Delta_{vec} &= \Delta \otimes (1, \ldots, K)^{T} ~~\text{with} ~~  \Delta \sim   \mathcal{N}_{N}(0_{N}, \tau_{T}^{2} Q_{D}^{-1})\\
    \Psi_{vec} &= \Phi_{vec} + \Delta_{vec},\\
 \psi_{tj} &= (\Psi_{vec})_{(t-1)\cdot K +j},\\
y_{tij} &= X_{tij}^{T}\beta + \psi_{tj} + r_{0ij} + r_{1ij}t + e_{tij}, ~ t=1, \ldots,N, ~~ i=1, \ldots n,~ j=1,\ldots, K,
    \end{array}
\right.
\end{equation}
where

$\displaystyle Q = \rho_{S} R + (1 - \rho_{S}) I$, $~ R_{jk} = -w_{jk}$,  $j \neq k$, $R_{jj} = \sum\limits_{k \neq j}w_{jk}$,  $~ j,k \in \{1, \ldots, K\}$, $Q_{D} = \rho_{T} R^{D} + (1 - \rho_{T}) I^{D}$, ~~ $R_{tl}^{D} = -d_{tl}$ for  $t \neq l$,  $R_{tt}^{D} = \sum\limits_{l \neq t}d_{tl}$, $ ~ t,l \in \{1, \ldots, N\}$. $I$ and $I^{D}$ are $K$ and $N$ dimensional unit matrices, respectively.
$\otimes$ denotes the Kronecker product.

\subsection{Simulation of the Spatial effect}
We use the following equation:

\begin{equation}\label{simulate_outcome_cs}
\left\{
  \begin{array}{ll}
    (\psi_{1}, \psi_{2},\ldots, \psi_{K})^{T} \sim  \mathcal{N}_{K}(0, \tau^{2} Q^{-1}),\\
   y_{ij} = X_{ij}^{T}\beta + \psi_{j} + e_{ij}, i=1,\ldots,N,~~ j=1, \ldots, K,
  \end{array}
\right.
\end{equation}

where
$I$ denotes a $K\times K$ unit matrix. $\tau^{2}$ is a variance parameter that controls the strength of the overall spatial structure.

\end{document}